\documentclass[letterpaper,10pt,twocolumn,superscriptaddress,aps,prl]{revtex4-2}
\usepackage{amsfonts,amssymb,amsmath,amsthm,graphicx}
\usepackage{url}
\usepackage{comment}
\usepackage[separate-uncertainty = true,multi-part-units = brackets]{siunitx}
\usepackage[colorlinks=true, citecolor=blue,urlcolor=blue,linkcolor=blue,filecolor=black]{hyperref}
\begin{document}


\title{Atomically-thin Single-photon Sources for Quantum Communication}

\author{Timm Gao}
\affiliation{Institut f\"ur Festk\"orperphysik, Technische Universit\"at Berlin, 10623 Berlin, Germany}
 
\author{Martin v. Helversen}
\affiliation{Institut f\"ur Festk\"orperphysik, Technische Universit\"at Berlin, 10623 Berlin, Germany}

\author{Carlos Anton-Solanas}
\affiliation{Institut für Physik, Carl von Ossietzky Universit\"at Oldenburg, 26111 Oldenburg, Germany}

\author{Christian Schneider}
\affiliation{Institut für Physik, Carl von Ossietzky Universit\"at Oldenburg, 26111 Oldenburg, Germany}

\author{Tobias Heindel}
\email[Corresponding author: ]{tobias.heindel@tu-berlin.de}
\affiliation{Institut f\"ur Festk\"orperphysik, Technische Universit\"at Berlin, 10623 Berlin, Germany}



\begin{abstract}
To date, quantum communication widely relies on attenuated lasers for secret key generation. In future quantum networks fundamental limitations resulting from their probabilistic photon distribution must be overcome by using deterministic quantum light sources. Confined excitons in monolayers of transition metal dichalcogenides (TMDCs) constitute an emerging type of emitter for quantum light generation. These atomically-thin solid-state sources show appealing prospects for large-scale and low-cost device integration, meeting the demands of quantum information technologies. Here, we pioneer the practical suitability of TMDC devices in quantum communication. We employ a $\mathrm{WSe}_2$ monolayer single-photon source to emulate the BB84 protocol in a quantum key distribution (QKD) setup and achieve click rates of up to \SI{66.95}{\kilo\hertz} and antibunching values down to $0.034$ - a performance competitive with QKD experiments using semiconductor quantum dots or color centers in diamond. Our work opens the route towards wider applications of quantum information technologies using TMDC single-photon sources.
\end{abstract}

\maketitle

\section{Introduction}
Quantum information technologies rely on the availability of high-performance sources emitting flying qubits \cite{Kimble2008}. Due to the challenges in the fabrication of practical quantum light sources emitting single- and entangled-photon states on-demand, implementations using faint lasers, or weak coherent pulses (WCPs), took the lead in the design and development of practical quantum communication systems. The invention of decoy-state protocols mitigated severe security loopholes related to the Poissonian photon statistics \cite{Wang2005, Ma2005}, but at the cost of additional protocol overhead. Solid-state quantum light sources \cite{Aharonovich2016}, on the other hand, offer deterministic operation in combination with engineered device functionalities, showing up prospects to overcome fundamental limitations of probabilistic sources. Semiconductor quantum dots \cite{Michler2000, Tomm2021} and color centers in diamond \cite{Hensen2015} are currently the main candidates among quantum emitters in the solid-state. However, monolayers of transition metal dichalcogenides (TMDCs) encounter increasing interest, due to the ease and flexibility in engineering their photonic properties and their straight-forward integration in photonic devices \cite{Mak2016, Luo2018, Iff2021, Turunen2022}. Single-photon emission from trapped excitons in atomically thin crystals was initially confirmed in monolayers of $\mathrm{WSe}_2$ \cite{Tonndorf:15,He2015,Srivastava2015,Chakraborty2015,Koperski2015} and later complemented using $\mathrm{WS}_2$ \cite{PalaciosBerraquero2016}, $\mathrm{MoS}_2$\cite{Barthelmi2020},  $\mathrm{MoSe}_2$\cite{Yu2021} and $\mathrm{MoTe}_2$ \cite{Zhao2021}. In addition, correlated pair emission from the biexciton-exciton radiative cascade was experimentally shown in $\mathrm{WSe}_2$\cite{He2016}, outlining the possibility to generate entangled-photon pairs. From a technological viewpoint, TMDC-based quantum light sources feature interesting advantages over other solid-state implementations: a relatively simple and low-cost fabrication, the tailoring of the excitonic properties via multi-stacking of monolayers \cite{Novoselov2016}, giant susceptibility to external strain \cite{Iff2019}, and a deterministic control on the spatial position of the quantum emitters \cite{Branny2017, PalaciosBerraquero2017}, even in optical cavities \cite{Iff2021}. In addition, TMDC-based single-photon sources cover all three telecom windows centered around \SI{850}{\nano\meter}, \SI{1300}{\nano\meter}, and \SI{1550}{\nano\meter} \cite{Zhao2021}, including wavelengths compatible with atomic transitions used for quantum memories \cite{Chou2007}. The first telecom window, addressed in this work, is particularly suited for free-space optical communication scenarios, including air-to-ground links using airplanes \cite{Nauerth2013} or satellites \cite{Liao2017}. Despite the immense progress seen in this field over the last decade \cite{Liu2019}, applications of TMDC quantum emitters in quantum information remained elusive so far.
Here we realize a bright single-photon source based on a strain engineered $\mathrm{WSe}_2$ monolayer and demonstrate the feasibility of quantum key distribution (QKD) via the BB84 protocol \cite{BB84}. We evaluate the single-photon purity, or $g^{(2)}(0)$, the quantum bit error ratio (QBER), as well as the secret key rates expected in full implementations of QKD and explore routines to optimize the achievable performance in terms of the tolerable losses. A comparative analysis reveals that our atomically-thin single-photon source is readily competitive with QKD experiments using semiconductor quantum dots and color centers in diamond.

\section{Results}
\label{sec:results}

\subsection*{Experimental QKD Setup}

\begin{figure*}
	\centering\includegraphics[]{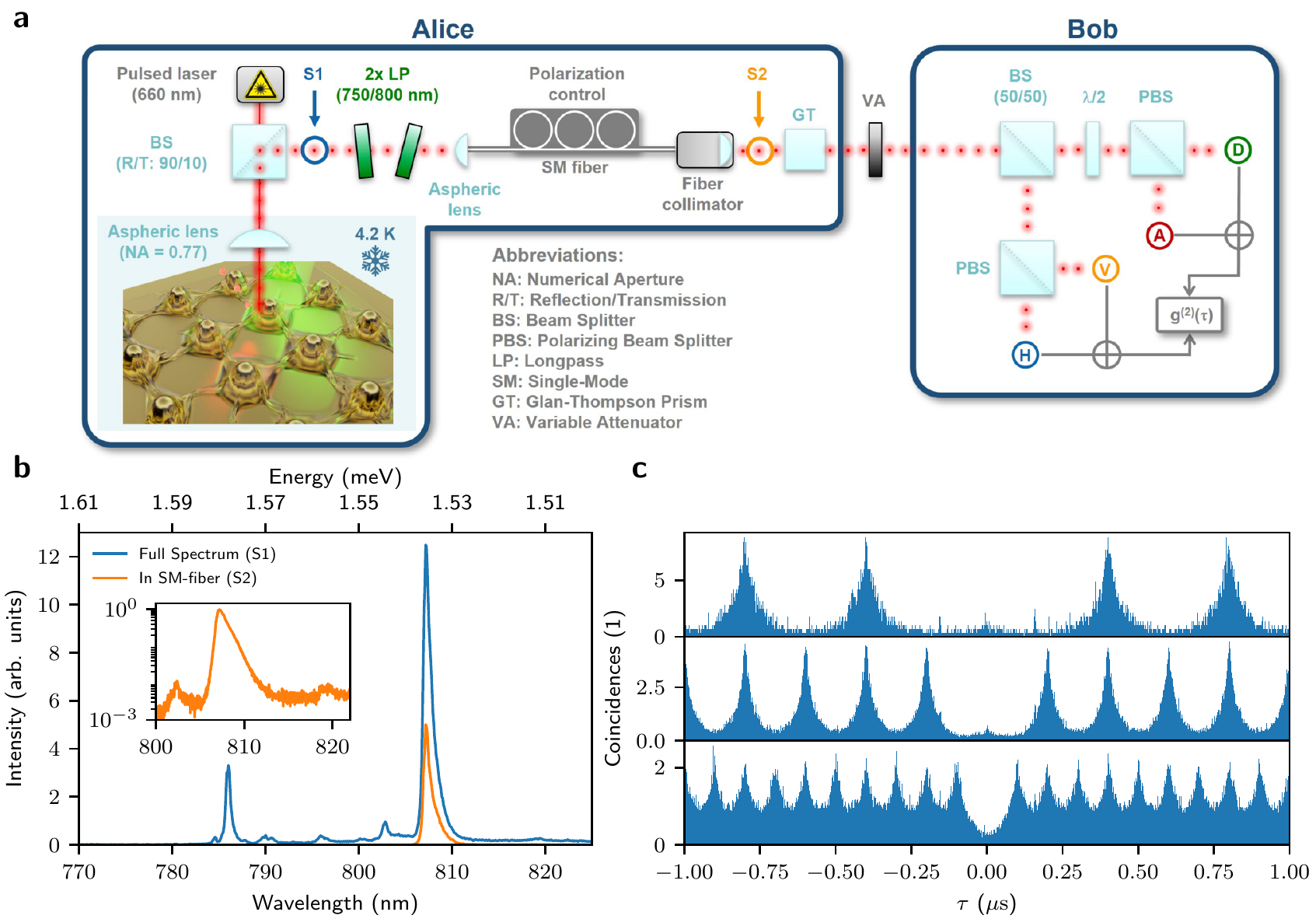}
	\caption{BB84-QKD setup employing an atomically thin single-photon source. \textbf{a} The transmitter (Alice) comprises a strain engineered $\mathrm{WSe}_2$ monolayer inside a closed cycle cryocooler, a diode-laser with variable repetition rate, dielectric filters, and an optical fiber together with a polarizer for adjusting the polarization states. Alice sends single-photon pulses with fixed polarization (H, V, D, A) in a back-to-back configuration. Channel losses are introduced by a variable attenuator. The receiver (Bob) comprises a four-state polarization analyzer based on bulk polarization optics. \textbf{b} Emission spectra of the $\mathrm{WSe}_2$ monolayer sample recorded at different locations in the setup: before (blue) and after (orange) spectral filtering and coupling to the SM fiber. The quantum emitter selected for the experiments in this work emits at a wavelength of \SI{807}{\nano\meter}. \textbf{c} Second-order autocorrelation histograms of the triggered quantum emitter at clock rates of \SI{2.5}{\mega\hertz}, \SI{5.0}{\mega\hertz}, and \SI{10.0}{\mega\hertz}. The autocorrelation is evaluated from the events registered at all four receiver channels.}
	\label{fig1}
\end{figure*}

The QKD setup used in our experiments is illustrated in Fig.~\ref{fig1}\,a. The transmitter (Alice) contains the single-photon source. For the source implementation, we followed a strategy detailed in ref.~\cite{Tripathi2018}, based on a strain engineered $\mathrm{WSe}_2$ monolayer device, containing localized quantum emitters (see Methods section \lq TMDC monolayer device\rq for details on the sample). The emission of a single emitter is collected in a confocal setup using spectral and spatial filtering via interference filters and the coupling to a single-mode (SM) optical fiber. The preparation of single-photons in four different polarization-states for the BB84 protocol is implemented in a fiber polarization controller in combination with a Glan-Thompson prism. The key parameters relevant for QKD are determined by preparing the polarization states in sequential measurement runs. The flying qubits are then sent to the receiver module (Bob) via a short free-space link, either in back-to-back configuration, i.e., without additional loss, or including a variable attenuator for emulating transmission losses. The receiver (Bob) comprises a four-state polarization decoder with passive basis choice and silicon-based single-photon detectors. Here, photons in the BB84-states are decoded in four different output channels (see Methods section \lq QKD Receiver\rq for details on the experimental setup).

\subsection*{QKD Performance}

To evaluate the performance of the source in our QKD setup, we select a localized $\mathrm{WSe}_2$ emitter featuring a single, bright emission line at \SI{807.3}{\nano\meter}.  Figure~\ref{fig1}\,b displays spectra of the selected emitter recorded before (blue) and after (orange) spectro-spatial filtering and coupling to the SM fiber under triggered optical excitation at \SI{5.0}{\mega\hertz}. Here, the observed asymmetric line shape (\SI{0.6}{\nano\meter} full width at half-maximum at saturation) most likely results from spectral diffusion \cite{Tripathi2018} in combination with efficient coupling to acoustic phonons \cite{He:16}. To evaluate the single photon purity of the flying qubits employed in the quantum channel of our QKD setup, we extract the second-order intensity autocorrelation $g^{(2)}(\tau)$ by correlating the detected photon events at all four output channels in the Bob module. The corresponding $g^{(2)}(\tau)$ histograms recorded at saturation of the quantum emitter are depicted in Fig.~\ref{fig1}\,c for different clock rates of \SI{2.5}{\mega\hertz}, \SI{5.0}{\mega\hertz}, and \SI{10.0}{\mega\hertz}. All three histograms clearly reveal a high single-photon purity with strongly suppressed coincidences around zero-time delay. The corresponding antibunching values $g^{(2)}(0)$, extracted by integrating the raw coincidences over a full repetition period, are 0.16 (\SI{2.5}{\mega\hertz}), 0.17 (\SI{5.0}{\mega\hertz}), and 0.25 (\SI{10.0}{\mega\hertz}) measured at saturation. Here, the noticeable increase in the integrated $g^{(2)}(0)$ at the highest clock rate arises from the overlap of the correlation peaks neighboring the one at zero delay (c.f., further below for a detailed discussion on limiting factors and a comparison to the literature). For the following studies, we choose a clock rate of \SI{5.0}{\mega\hertz} as the optimal trade-off for our experiments. 

\begin{figure*}
	\centering\includegraphics[]{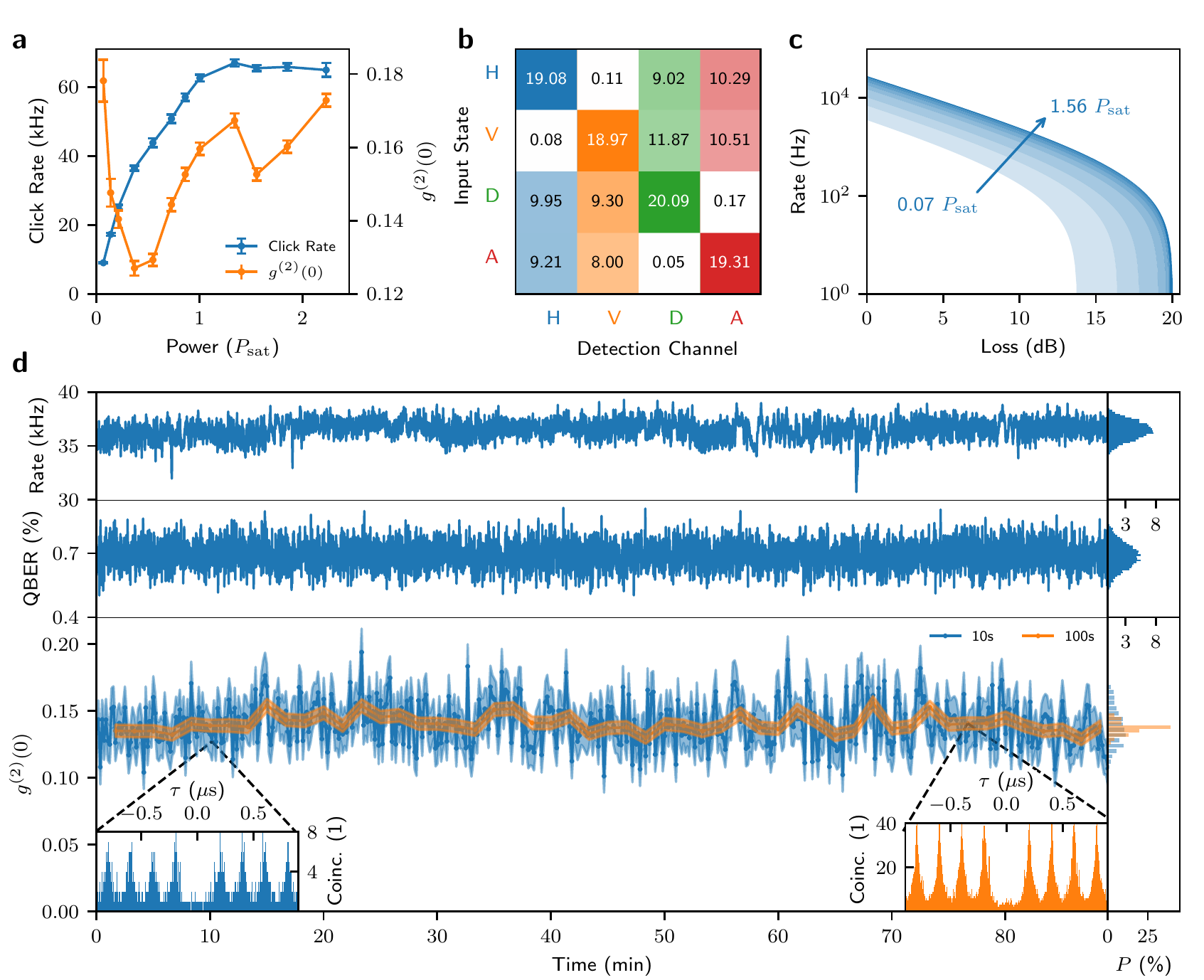}
	\caption{Performance evaluation of the $\mathrm{WSe}_2$ single-photon source device for applications in QKD. \textbf{a} Excitation power resolved series of the click rate and $g^{(2)}(0)$. The source reaches saturation at \SI{52.5}{\micro\watt}. Solid lines are a guide to the eye. Error bars are the standard deviations. \textbf{b} Input-Output characterization of Bob for extraction of the QBER. The detection-rates in kHz observed at each channel are stated and additionally encoded in the field's opacity. \textbf{c} Rate-loss diagram considering the experimental data from a. Although the $g^{(2)}(0)$ is not at the minimum at saturation, the achievable distance and secret key rate for each distance has its maximum at saturation. \textbf{d} Real-time security monitoring over exemplary time of \SI{1.5}{\hour} ($P = 0.43 P_{\mathrm{sat}}$, H input polarization). The QBER is calculated from all registered events. In full implementations using only subsets for an estimation, this value might be higher. Antibunching $g^{(2)}(0)$ of the single-photon source evaluated by correlating all registered events in non-overlapping blocks for \SI{10}{\second} (blue) and \SI{100}{\second} (orange) accumulation time (exemplary $g^{(2)}(\tau)$ histograms shown as insets). See Supplementary Figure~6 for details on the accumulation time.}
	\label{fig2}
\end{figure*}

Next, we investigate how the optical pumping conditions of the quantum emitter influence the performance in QKD. Figure~\ref{fig2}\,a shows the click rate (total signal from all four channels) detected at the receiver in back-to-back configuration, i.e. negligible losses in the quantum channel, together with the corresponding $g^{(2)}(0)$ of the single-photon source as a function of the pumping strength. In this experiment, we utilize the receiver module itself for characterizing our source. Note, that in full implementations Alice must monitor the source also on her side. This can be implemented by using a tap-coupler or a polarizing beam splitter in front of the encoding electro-optical modulator, which splits a fraction of the single-photon pulses for monitoring the source during transmission. Without the in-situ monitoring of the source, security loopholes can arise, e.g. due to underestimation of multi-photon events. We observe an almost monotonic increase in the click rate up to \SI{66.95\pm1.07}{\kilo\hertz} at saturation pump power ($P_{\mathrm{sat}}$) corresponding to a brightness of the single-photon source under pulsed excitation of \SI{2.9\pm0.3}{\percent} in the SM fiber. Under continuous wave excitation we observe a maximum count rate of \SI{530}{\kilo\hertz}, corresponding to a single-photon flux of \SI{1.15\pm0.23}{\mega\hertz} in the SM fiber (see Methods section \lq Estimating the source brightness and spectral background contribution\rq for details on the efficiency estimate). For QKD applications, the important parameter to extract is the mean photon number per pulse into the quantum channel $\mu$. With the transmission $\eta_{\mathrm{Bob}}=0.56$ of the receiver module including the transmission of the optics and detector efficiencies, we achieve maximum values of up to $\mu = 0.024$. In a full implementation including an electro-optical modulator, with typical insertion losses of \SI{3}{\decibel} for fiber-based and \SI{1}{\decibel} for free-space implementations, the overall losses will be slightly higher resulting in lower values of $\mu$.
The antibunching value shows a more complex dependence. Under weak pumping, $g^{(2)}(0)$ first decreases with increasing excitation power to a minimum value of $0.127\pm0.002$ at $0.37~P_{\mathrm{sat}}$. Further increasing the pumping strength, $g^{(2)}(0)$ shows a moderate increase up to $0.172$. While the decrease of $g^{(2)}(0)$ at weak pumping is attributed to an improved signal-to-noise ratio, its increase at stronger pumping results from noticeable spectral contributions of residual emission lines close to the selected emitter (cf. Fig.~\ref{fig1}\,b, inset), which are not perfectly rejected by the spectro-spatial filter used on Alice's side. The additional dip at $1.5~P_{\mathrm{sat}}$ might indicate different decay paths or pumping into different energy levels of the localized quantum emitter. The contribution of the spectral background to $g^{(2)}(0)$ is estimated to $0.040$ at $0.37~P_{\mathrm{sat}}$, partly explaining the deviation from an ideal single-photon source (see ref.~\cite{Brouri:00} and see Methods section \lq Estimating the source brightness and spectral background contribution\rq for details). Using temporal filtering as a post process, $g^{(2)}(0)$ can be reduced down to $0.034\pm0.002$ for an acceptance time window width of \SI{24}{\nano\second} (see Supplementary Figure~4). This value compares favorably with the lowest antibunching values of $g^{(2)}(0) = 0.058\pm0.003$ reported to date for TMDC-based quantum emitters under pulsed optical excitation at comparable wavelength \cite{Zhao2021}. In the following, we intentionally neither apply background corrections nor fitting routines to the $g^{(2)}(0)$ values; this is important for QKD applications, where both signal and background photons contribute to the overall key. An underestimation of the $g^{(2)}(0)$ leads to an underestimation of the multi-photon pulses in the quantum channel (see also Methods section \lq Key rate calculations\rq for details) which in turn results in an information leak to an attacker. Even if using temporal filtering to optimize the performance of our QKD setup further below, we use the full $g^{(2)}(0)$ value as worst-case assumption. To exploit the positive effect of temporal filtering on the $g^{(2)}(0)$, Alice has to apply the temporal filter already on her side, e.g. by implementing an additional electro-optical intensity modulator before the quantum channel (see ref.~\cite{Kupko2020} for a detailed discussion). This reduces the multi-photon pulses in the quantum channel and makes them no longer accessible to an attacker. For our setup, the additional losses of a modulator would not be compensated by the reduction of multi-photon pulses (see Supplementary Figure~5). 

To estimate the QBER achievable in our QKD setup, the fiber-coupled single-photon pulses are sequentially prepared using a static linear polarizer (see Methods section \lq QKD transmitter\rq) set to horizontal (H), vertical (V), diagonal (D), or antidiagonal (A) polarization. The pulses are recorded at the receivers's four detection channels. The resulting experimental data are summarized in Fig.~\ref{fig2}\,b in a 4x4 matrix with each entry corresponding to the event-rate in the respective measurement configuration (see Supplementary Figure~1 for the full time-resolved data underlying this illustration). The working principal of the polarization decoder results in prominent diagonal elements, i.e., for a given input-state, almost all photons are detected in the respective channel of the corresponding basis, whereas a probabilistic projection is observed in the conjugate basis. Erroneous detection events at the wrong channel within one basis, e.g., events detected in the V-channel for H-input, are much less probable and cause the finite QBER to be considered in the security analysis in the following. In our QKD analysis, mainly two sources contribute to the errors: Background events resulting from residual stray light plus dark counts of the four detectors (\SI{80}{\hertz} in total) and optical imperfections inside the receiver. Additional deviations from an ideal setup are detection efficiency mismatches between channels, which need to be considered in full implementations \cite{Lydersen2010}. From the experimental data in Fig.~\ref{fig2}\,b we extract a QBER of only \SI{0.52}{\percent}.This value gives a lower bound for full implementations as it describes the limit the receiver optics set to the overall QBER. In full QKD implementations, additional errors are expected due to imperfections in the polarization preparation using electro-optical modulators. These are typically in the range of \SI{1}{\percent}. In the following, we apply the lower bound found in our characterization.
Based on the previous parameter analysis, we can extrapolate the secret key rate expected in full implementations of QKD as a function of tolerable losses. To this end we evaluate the secret key rate $S_{\infty}$ in the asymptotic limit, i.e., assuming an infinitely long key, by following the formalism presented in ref.~\cite{Gottesman2004} using an upper bound for the multi-photon contribution as in ref.~\cite{Waks2002a} (cf. Methods section \lq Key rate calculation\rq for details). Note, that in full implementations a finite state preparation quality needs to be considered as discussed in ref.~\cite{Tomamichel2012}. While being negligible in our QKD testbed (due to the use of a linear polarizer), this effect becomes relevant in the case of a dynamic state preparation using electro-optical modulators, which limit the state preparation quality (see Supplementary Figure~2 and corresponding discussions for details on the effects of using electro-optical modulators). The resulting rate-loss dependencies of our QKD experiment are displayed in Fig.~\ref{fig2}\,c as a function of the pumping strength. While each curve follows a linear trend (in logarithmic scaling) at low to moderate losses, a multi-exponential drop is observed in the high-loss regime determining the well-known distance limit in point-to-point quantum communication \cite{Takeoka2014, Pirandola2017}. At a low excitation power of $0.05~P_{\mathrm{sat}}$ the maximally achievable tolerable loss is limited to \SI{13.82}{\decibel}. Although $g^{(2)}(0)$ increases at strong pumping (cf. Fig.~\ref{fig2}\,a), we find the largest secret key rate and maximally achievable tolerable loss of \SI{20.14}{\decibel} inside the transmission link to be expected at saturation of our quantum emitter (see Fig.~\ref{fig2}\,c). Note, that in our QKD testbed the increase in $g^{(2)}(0)$, which in turn results in a larger multi-photon emission probability, is overcompensated by an increase in $\mu$, resulting in larger detection rates at higher pump strength.
For practical implementations in quantum information, the temporal stability of the employed single-photon source is an important characteristic and many QKD protocols benefit from monitoring the security parameters in real-time to certify its security. Time-traces of the key parameters of our fiber-coupled atomically-thin single-photon source, recorded during a measurement period of \SI{1.5}{\hour}, are presented in Fig.~\ref{fig2}\,d. Here, the click rate, the QBER, and the $g^{(2)}(0)$ value are depicted together with their corresponding probability distribution (cf. right panel). Photon flux and QBER are stable over time with average values of \SI{36.5\pm0.9}{\kilo\hertz} and \SI{0.69\pm0.06}{\percent}, respectively. The same holds true for the antibunching with an average value of $g^{(2)}(0) = 0.14$, shown for two different accumulation times of \SI{10}{\second} (blue) and \SI{100}{\second} (orange) resulting in standard deviations of $0.015$ and $0.006$, respectively (see Supplementary Figure~6 for details on the accumulation time). In the following section, the performance of our system will be optimized and compared with other technologies (see Tab.~\ref{tab:1}). 

\section*{Optimization and Benchmarking}

\begin{figure*}
	\centering\includegraphics[]{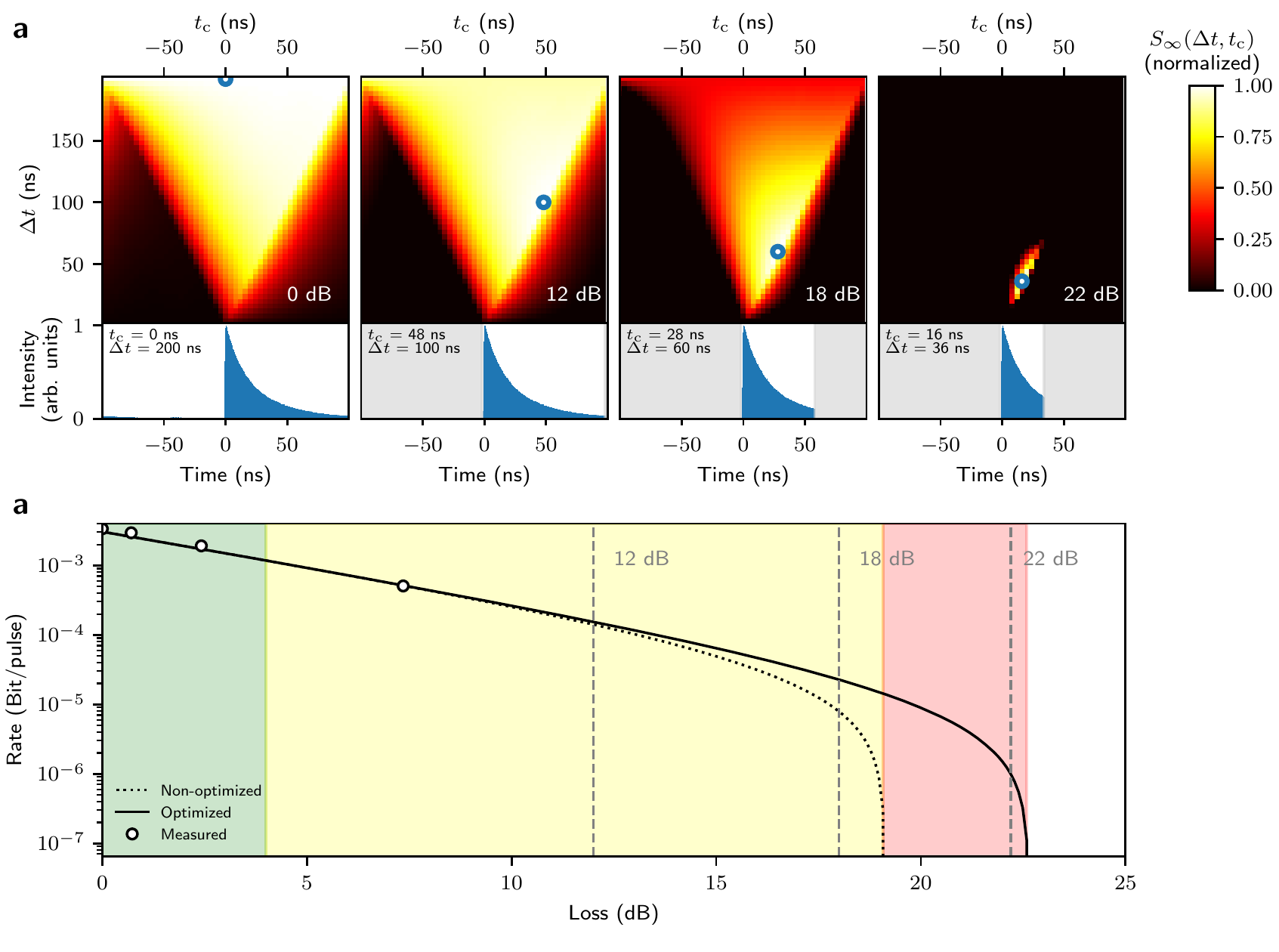}
	\caption{Key rate optimization via 2D temporal filtering. \textbf{a} Expected secret key rate fraction $S_{\infty}(\Delta t, t_{{\mathrm{c}}})$ as a function of the acceptance time window $\Delta t$ and $t_{{\mathrm{c}}}$. From the experimental results in back-to-back configuration (\SI{0}{\decibel} loss in quantum channel) at $0.43~P_{\mathrm{sat}}$ and fixed $g^{(2)}(0) = 0.134$, three different loss scenarios are simulated. Blue circles mark the optimal parameter sets resulting in a maximal secure key rate. \textbf{b} Rate loss diagram comparing the non-optimized secret key rate (compare Fig.~\ref{fig2}\,c) and optimized case. For each individual loss, the secret key rate was optimized according to a. Open circles correspond to experimental results using a variable attenuator in the quantum channel.}
	\label{fig3}
\end{figure*}

For the operation of practical QKD systems, the optimization of parameters entering the secure key rate (e.g., intensities, acceptance time windows) is a crucial step in gaining optimal performance for a given link scenario \cite{Wang2019}. This becomes important in communication scenarios which are affected by time-dependent variations, as e.g., the link transmission in free-space optical settings between stationary \cite{Ursin2007} or moving platforms \cite{Nauerth2013, Liao2017}. Here, the transmission losses inside the quantum channel can drastically vary due to the weather conditions or the transmitter-receiver distance. 
In the following, we explore temporal filtering, to optimize the performance of our single-photon QKD setup. As mentioned in the previous section, detector dark counts as well as imperfections of polarization optics contribute to the overall QBER. While the detector noise is uncorrelated in time, erroneous detection events due to optical imperfections are correlated with the clock-rate of our experiment. Hence, the contribution of detector noise can be efficiently reduced by narrowing the temporal acceptance window on the receiver side, resulting in lower QBERs - ideally down to the limit of $\mathrm{QBER} = \SI{0.42}{\percent}$ of our present setup. As the applied temporal filter will also reduce the detection events available for key distillation, there is an optimal trade-off between QBER and sifted key for which the secure key can be maximized for a given channel loss (see Supplementary Figure~3 for details). Due to the asymmetric temporal shape of the single-photon pulses, it is beneficial to vary both the width $\Delta t$ and the center $t_{\mathrm{c}}$ of the acceptance time window as seen in Fig.~\ref{fig3}\,a. Here, the asymptotic secret key rate $S_{\infty}(\Delta t, t_{\mathrm{c}})$, simulated from the experimental data in Fig.~\ref{fig2}\,d, is depicted in heat-maps corresponding to four different loss scenarios in the quantum channel of \SI{0}{\decibel}, \SI{12}{\decibel}, \SI{18}{\decibel}, and \SI{22}{\decibel} (see Methods section \lq Parameter optimization\rq for details). Note, that the loss budget stated here is only the contribution of the transmission channel to the overall losses, i.e., the loss from the last optical element on Alice’ side to the first trusted optical element on Bob’s side. The loss of the receiver module is added for during the calculation of the secret key rate. The optimal parameter settings for the temporal acceptance window, resulting in the largest possible secret key rate, are marked by blue circles in the heat-maps together with the resulting single-photon pulse in the respective lower panels. While the gain achieved via the 2D temporal filtering is small in the low-loss regime, the improved signal-to-noise ratio results in clear advantages at higher losses. This is clearly seen from the extrapolated rate-loss dependencies in Fig.~\ref{fig3}\,b, comparing the non-optimized (dotted line) with the optimized (solid line) case, with the secret key rate $S_{\infty}(\Delta t, t_{{\mathrm{c}}})$ being optimized for each individual loss. Three regimes can be distinguished: While in the low-loss regime (green), the gain achieved via temporal filtering is marginal, the optimization leads to an improved $S_{\infty}(\Delta t, t_{{\mathrm{c}}})$ in the intermediate-loss regime (yellow) between \SI{4}{\decibel} and \SI{19.08}{\decibel}, where the upper limit corresponds to the maximally tolerable loss in the non-optimized case. In the high-loss regime (red), secure communication is only possible if a parameter optimization is applied. Overall, this routine allows us to extend the maximally tolerable transmission loss by \SI{3.51}{\decibel} to \SI{22.59}{\decibel} using the optimal settings for the temporal filter width $\Delta t$ and center $t_{{\mathrm{c}}}$. Assuming a free-space optical communication link at clear-sky weather conditions (\SI{0.06}{\decibel/\kilo\meter}-\SI{0.08}{\decibel/\kilo\meter} atmospheric loss at \SI{785}{\nano\meter} \cite{Ursin2007}, neglecting beam spread), this corresponds to an extension of the achievable communication distance by \SI{43.9}{\kilo\meter}.

\begin{table}
	\centering
	\begin{tabular}{|l|l|l|l|l|l|}
		\hline
		& ref.~\cite{Waks2002} & ref.~\cite{Leifgen_2014} &  ref.~\cite{Takemoto2015} & This work & Improved \\
		\hline
		$\lambda$ (nm) & $877$ & $600-800$ & $1580.5$ & $807$ & $807$ \\
		\hline
		$\mu$ & $0.007$ & $0.029$ & $0.009$ & $0.013$ & $0.050$ \\
		\hline
		$p_{\mathrm{dc}}$ & $1.05e-6$ & $24.00e-6$ & $0.3e-6$ & $16.00e-6$ & $2.0e-6$ \\
		\hline
		$e_{\mathrm{detector}}$ & $0.025$ & $0.030$ & $0.023$ & $0.008$ & $0.008$ \\
		\hline
		$\eta_{\mathrm{Bob}}$ & $0.24$ & $0.31$ & $0.048$ & $0.56$ & $0.56$ \\
		\hline
		$g^{(2)}(0)$ & $0.14$ & $0.09$ & $0.0051$ & $0.133$ & $0.05$ \\
		\hline
	\end{tabular}
	\caption{Parameter sets for describing the performance of the state-of-the-art SPS-based QKD free-space implementations for quantum dots \cite{Waks2002}, color centers in diamond \cite{Leifgen_2014}, fiber-based implementation for quantum dots \cite{Takemoto2015} and this work. Here, $p_{\mathrm{dc}}$ is the dark count probability and $e_{\mathrm{detector}}$ the static contribution of the receiver module to the QBER.}
	\label{tab:1}
\end{table}

\begin{figure}
	\centering\includegraphics[]{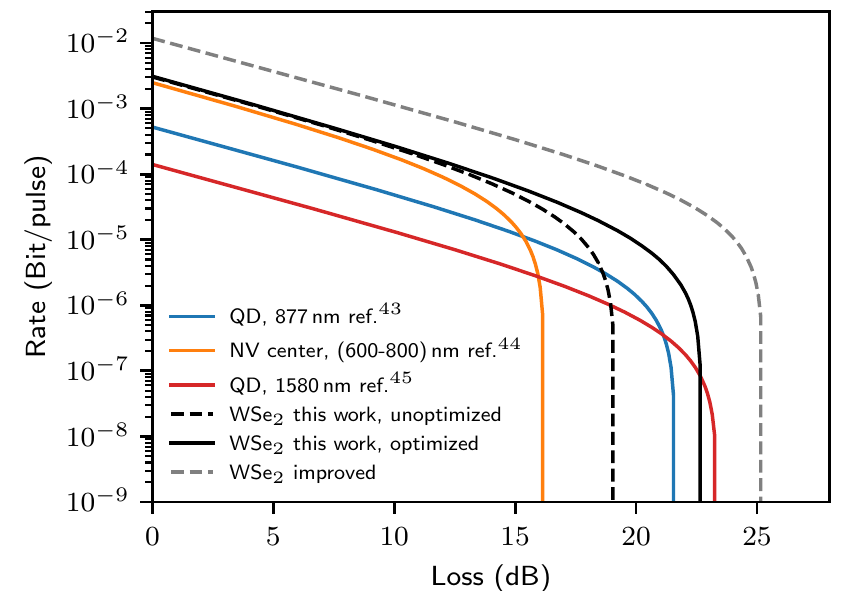}
	\caption{Benchmarking of our $\mathrm{WSe}_2$ single-photon source against previous QKD experiments. Rate-loss diagram considering our experimental data from Tab.~\ref{tab:1} and the parameters from previous QKD experiments using semiconductor quantum dots (QD) \cite{Waks2002}, \cite{Takemoto2015}, and Nitrogen vacancy (NV) centers in diamond \cite{Leifgen_2014} (see Methods section \lq Parameter optimization\rq for details on the key rate calculation). The black curves show the expected performance of our $\mathrm{WSe}_2$-based source without (dashed line) and with (full line) optimization via temporal filtering. Implementing moderate improvements to our source (grey dashed line), tolerable losses exceeding \SI{25}{\decibel} become possible, bringing satellite links within reach \cite{Liao2017}.}
	\label{fig4}
\end{figure}

To put our results into perspective, we compare our results with previous QKD experiments using non-classical light sources. As summarized in a recent review article \cite{Vajner2022}, several proof-of-principle experiments using semiconductor quantum dots for single-photon (e.g., refs.~\cite{Waks2002,Intallura_2009,Heindel_2012,Takemoto2015}) or entanglement-based \cite{Basset2021, Schimpf2021} QKD as well as color centers in diamond for single-photon QKD (e.g., refs.~\cite{Beveratos2002, Leifgen_2014}) have been reported to date, covering all three telecom windows and different levels of device integration. Recently, also quantum emitters in hexagonal boron nitride (hBN) \cite{Zeng2022} as well as molecules of polyaromatic hydrocarbons \cite{Murtaza2022} were considered and evaluated for their application in QKD, including an implementation of the B92 protocol \cite{Bennett1992a} using a hBN-based SPS \cite{Samaner2022}. For our following comparison, we restrict ourselves to the state-of-the-art of BB84 QKD: the pioneering work by Waks et al. \cite{Waks2002}, which reported the largest secure key rate for quantum dot based QKD to date, Leifgen et al. \cite{Leifgen_2014}, evaluating nitrogen- and silicon-vacancy centers in diamond for QKD, and Takemoto et al. \cite{Takemoto2015} with the longest distance achieved for SPS-based QKD so far (c.f. Tab.~\ref{tab:1} for the parameter sets). The rate-loss dependencies of these reports together with our present work are presented in Fig.~\ref{fig4}. The single-photon QKD experiment using quantum dots by Waks et al. \cite{Waks2002} realized at an operation wavelength of \SI{877}{\nano\meter} reported the largest secure key rate to date, the experiment of Takemoto et al. \cite{Takemoto2015} reports the highest tolerable losses. The work of Leifgen et al. \cite{Leifgen_2014} represents the current state-of-the art for QKD implementations with color centers in diamond. The comparison with our present work reveals that our simple $\mathrm{WSe}_2$-based single-photon source is already competitive with state-of-the-art QKD experiments in terms of the expected secure bits per pulse and, if optimization routines are applied, also in terms of the tolerable losses, which exceed the values achieved in ref.~\cite{Waks2002} and are close to the once in ref.~\cite{Takemoto2015}. The mean photon number per pulse inside the quantum channel even clearly surpasses the performance of previous single-photon QKD experiments, which we attribute mainly to the use of a more efficient spectral filtering of the single photons. Straightforward improvements include the integration of TMDC monolayers in microcavities \cite{Luo2018, Iff2021} to reduce the radiative lifetime and increase the photon extraction efficiency. While the Purcell enhancement allows for smaller duty cycles, resulting in improved signal-to-noise ratios, the increased extraction efficiency directly translates to higher values of $\mu$. Both effects combined will lead to significant improvements in the achievable tolerable losses. Implementing already moderate improvements in the performance of the TMDC-based single-photon source (see Fig.~\ref{fig4}, dashed grey line: $\mu = 0.05$, $g^{(2)}(0) = 0.05$, and \SI{10}{\hertz} dark count rate), we anticipate maximally tolerable losses in the quantum channel exceeding \SI{25}{\decibel} to become possible. This loss budget brings free-space optical links between the Canary Islands \cite{Ursin2007} or even satellite-to-ground QKD within reach \cite{Liao2017}.

\section{Discussion}
We demonstrated the feasibility of quantum communication using an atomically-thin single-photon source based on a strain-engineered $\mathrm{WSe}_2$ monolayer. Implemented in a QKD setup emulating the BB84-protocol, the atomically-thin TMDC single-photon source shows a performance superior to previous QKD experiments using solid-state non-classical light sources, opening the route for low-cost large-scale applications in quantum information. Utilizing directly fiber-pigtailed TMDC-devices in combination with compact Stirling cryocoolers, user-friendly plug\&play quantum light sources will be developed, following our recent demonstration for QD-based devices \cite{Gao2022}. Further advances in material engineering might even bring room temperature operation within reach \cite{Luo2019}. Future real-world QKD-applications using highly integrated TMDC-based single-photon sources, will employ fast electro-optical modulator fed by quantum random number generators for polarization-state preparation, enabling secure communication between distant and moving platforms via free-space-optical links. In this context, also finite key-size effects need to be taken into account - a task for which recent work in the field promises substantial improvements over previous work \cite{Morrison2022}. Considering advanced implementations in quantum communication but also photonic quantum computing, important milestones to achieve will be the demonstration of high photon-indistinguishabilities of single and between multiple remote TMDC single-photon sources \cite{Turunen2022}.

\section{Methods}
\subsection{TMDC monolayer device}
The TMDC sample used in this work comprises a strain-engineered monolayer of $\mathrm{WSe}_2$ providing localized quantum emitters for single-photon generation. The sample is fabricated by mechanical exfoliation of sheets of $\mathrm{WSe}_2$ transferred to a nano-structured metallic surface, resulting from the deposition of \SI{200}{\nano\meter} of silver on a \SI{600}{\micro\meter} thick sapphire substrate capped with \SI{10}{\nano\meter} of chromium. The resulting surface contains silver nanoparticles of varying size, which induce wrinkles in the overlying $\mathrm{WSe}_2$ acting as strain centers for the excitonic emission. For a detailed characterization of this type of sample we refer the interested reader to ref.~\cite{Tripathi2018}.

\subsection{QKD transmitter}
At the heart of the transmitter (Alice), the strain engineered $\mathrm{WSe}_2$ monolayer device is mounted inside a closed-cycle cryocooler (attoDRY800 by Attocube Systems AG) cooled down to \SI{4.2}{\kelvin} including an aspheric lens for collecting the emission of single quantum emitters with a numerical aperture of $0.77$. The quantum emitters are optically triggered using a SM fiber-coupled pulsed diode laser (LDH-P-650 by PicoQuant GmbH) with variable repetition rate emitting at \SI{660}{\nano\meter}, which is directed to the sample via a 90:10 beamsplitter. The \SI{90}{\percent} port of the latter is used to collect the sample luminescence. Spectral filtering of a single emission line of a quantum emitter at the long-wavelength tale of the inhomogeneously broadened ensemble is achieved by two long-pass (LP) filters (cut-on wavelengths: \SI{750}{\nano\meter} and \SI{800}{\nano\meter}), before the emission is coupled to a single-mode (SM) optical fiber (type 780HP, \SI{5}{\micro\meter} core diameter, $\mathrm{NA} = 0.13$) with an aspheric lens (\SI{18}{\milli\meter} focal length). Noteworthy, here the use of LP filters (instead of a grating spectrometer) in combination with the additional spatial filtering via the SM fiber enables us to achieve a low-loss optical setup. For preparing single-photon pulses in the four different polarization-states required for emulating the BB84 protocol, a fiber-based polarization controller in combination with a Glan-Thompson prism is used. Note, while full implementations of QKD require a dynamic modulation of the polarization states (e.g., via electro-optical modulators), we statically prepare the polarization states in sequential measurement runs in the QKD system used in this proof-of-concept work. This allows us to determine all key-parameters for QKD and their limits.

\subsection{QKD receiver} 
The polarization qubits prepared at the transmitter are then sent to the receiver module (Bob) via a short free-space link, either in back-to-back configuration, i.e., without additional loss, or including a variable attenuator (absorptive neutral density filters) for emulating transmission losses. Bob comprises a four-state polarization decoder with passive basis choice designed for operation in the wavelength range \SI{720}{\nano\meter}-\SI{980}{\nano\meter}. The measurement bases are chosen by a nonpolarizing 50:50 beamsplitter cube. The final discrimination of the polarization state is realized with polarizing beamsplitters, one of them has an additional half-wave plate in front to project incoming photons from the diagonal/antidiagonal basis to the beamsplitter axes. Each of the four output ports has a fiber collimator with attached optical multimode fiber (FG050LGA, \SI{15}{\meter}) connected with single-photon counting modules (COUNT-T100-FC, Laser Components GmbH) having a mean efficiency of \SI{80}{\percent} at \SI{810}{\nano\meter}, \SI{500}{\pico\second} timing resolution and \SI{20}{\hertz} dark count rate. The stream of detection events is registered and digitized by a time-to-digital converter (quTAG, qutools GmbH) and synchronized with the excitation laser. More details can be found in ref.~\cite{Kupko2020}. Characterizing the polarization decoder with our single-photon source, we determine the QBER for each detection channel individually ($\mathrm{QBER}_{\mathrm{H}} = \SI{0.57}{\percent}$, $\mathrm{QBER}_{\mathrm{V}} = \SI{0.42}{\percent}$, $\mathrm{QBER}_{\mathrm{D}} = \SI{0.84}{\percent}$, and $\mathrm{QBER}_{\mathrm{A}} = \SI{0.26}{\percent}$) resulting in an average QBER of \SI{0.52}{\percent}. For the key rate calculations, we assumed the worst case of $\mathrm{QBER}_{\mathrm{D}} = \SI{0.84}{\percent}$.

\subsection{Estimating the source brightness and spectral background contribution}
To estimate the single-photon flux our $\mathrm{WSe}_2$ emits into the SM fiber on Alice's side, we analyzed the transmission of our QKD setup. The efficiency of the receiver module Bob including the detector efficiencies and optics' transmission is \SI{56\pm5}{\percent} with additional losses in the polarizer, fiber, and mating sleeves. This results in a transmission of our setup of \SI{46\pm9}{\percent} from the SM fiber to the detectors. The total detection rate of up to \SI{66.95\pm1.07}{\kilo\hertz} thus corresponds to a single-photon flux in SM fiber of \SI{146\pm29}{\kilo\hertz} or a brightness of the SM fiber-coupled single-photon source of \SI{2.9\pm0.3}{\percent} under pulsed excitation at \SI{5}{\mega\hertz}. Under continuous wave excitation we observe a maximum count rate of \SI{530}{\kilo\hertz} at saturation of the quantum emitter corresponding to \SI{1.15\pm0.23}{\mega\hertz} in the SM fiber. To explore limiting effects to the single-photon purity of our source, we account for residual background contributions in the observed spectra. Assuming an uncorrelated background, the expected auto-correlation function can be expressed via $g_b^{\left(2\right)}(0)=1+\rho^2\left[g_{\mathrm{emitter}}^{(2)}\left(0\right)-1\right]$, where $\rho=S/\left(S+B\right)$ is the ratio of signal $S$ to background $B$ and $g_{\mathrm{emitter}}^{(2)}\left(0\right)$ is the single-photon purity of the emitter itself \cite{Brouri:00}. Extracting $\rho = 0.97$ from the inset in Fig.~\ref{fig1}\,b and assuming $g_{\mathrm{emitter}}^{(2)}\left(0\right) = 0$, results in a limit of $g_b^{\left(2\right)}\left(0\right)=0.040\pm0.003$ solely due to the spectral background, partly explaining the nonideal single-photon purity. The estimate of the spectral background is in good agreement with the temporally filtered value of $g^{(2)}\left(0\right)=0.034\pm0.002$, where residual laser emission is rejected.

\subsection{Key rate calculations}
Using the formalism presented in the ref.~\cite{Gottesman2004} the so-called GLLP rate, named after the authors Gottesman, Lo, L\"utkenhaus, and Preskill, reads
\begin{equation}
	S_\infty=S_{\mathrm{sift}}\left[A\left(1-h\left(e/A\right)\right)-f_{\mathrm{EC}}h\left(e\right)\right]. 
\end{equation}
Here, $S_{\mathrm{sift}}$ is the sifted key rate and $A=\left(p_{\mathrm{click}}-p_\mathrm{m}\right)/p_{\mathrm{click}}$ is the single-photon detection probability. The latter is a function of the multiphoton emission probability $p_\mathrm{m}$ and overall click probability $p_{\mathrm{click}} \approx \mu\cdot T\cdot\eta_{\mathrm{Bob}}+p_{\mathrm{dc}}$, with the channel transmission $T$, the transmission of the Bob module $\eta_{\mathrm{Bob}}$ including detector efficiencies, and the dark count probability $p_{\mathrm{dc}}$. The QBER is denoted with $e$, $h\left(e\right)$ the binary Shannon-entropy, and $f_{\mathrm{EC}}$ is the error correction efficiency. The factor $A$ is a correction for the finite $p_\mathrm{m}$ of practical quantum light sources. One possible way to estimate $p_\mathrm{m}$ in an experiment is the second order autocorrelation giving an upper bound of $p_\mathrm{m}=\mu^2g^{(2)}(0)/2$ \cite{Waks2002a}. The factor $\mu$ stands for the mean photon number per pulse inside the quantum channel and corresponds to the overall efficiency of Alice in the case of sub-Poissonian light sources. From the click rates measured at our receiver in back-to-back configuration and the transmission of Bob $\eta_{\mathrm{Bob}} = 0.56$, we determine the mean photon number per pulse into the quantum channel $\mu$ in our experiment. From $\mu$ and $g^{(2)}(0)$ it is then straightforward to extract $p_\mathrm{m}$. While we use neutral density filters to measure the parameters for estimating the secret key rate in Fig.~\ref{fig3}\,b, the simulated rate-loss graphs in Fig.~\ref{fig2}\,c and Fig.~\ref{fig3}\,b are obtained by using the dependence of $p_{\mathrm{click}}$ on the channel transmission.

\subsection{Parameter optimization}
Using the formalism presented in ref.~\cite{Kupko2020} we optimized the acceptance time window of detection events. By reducing the acceptance time window, the dark count probability $p_{\mathrm{dc}}$ decreases proportionally. At the same time, the number of accepted detection events and hence the click rate is reduced. The reduction of the available key material does not shrink proportionally to the acceptance window size $\Delta t$ but follows the arrival time probability distribution of the emitter. This distribution is asymmetrical due to the slow exponential decay of the spontaneous emission compared to the fast excitation. It is not enough to vary the acceptance window size $\Delta t$ around the maximum of the distribution. By additionally varying the center $t_{\mathrm{c}}$ one can cover all possible acceptance time windows. In Fig.~\ref{fig3}\,a we used a subset of \SI{136}{\second} (sifted block of \SI{5}{\mega bit}) of the measurement in Fig.~\ref{fig2}\,d and evaluated the timestamps for each acceptance time window individually to obtain a 2D heat-map depicting the (normalized) secret key rate depending on $\Delta t$ and $t_{\mathrm{c}}$. We use $50$ by $50$ equal steps (minimum $\Delta t = \SI{4}{\nano\second}$). The QBER and sifted key faction are varied,  $g^{(2)}(0) = 0.134$ is fixed to its unfiltered value. Finally, we simulate the rate-loss dependency using the parameter $p_{\mathrm{click}}$ as mentioned above. In this fashion the optimized rate loss graph in Fig.~\ref{fig4} is simulated with 2D optimization for each loss regime individually. 

\section{Data availability}
The data that support the plots within this paper and other findings of this study are available from the corresponding author upon reasonable request.

\section{Acknowledgments}
We acknowledge financial support by the German Federal Ministry of Education and Research (BMBF) via the project \lq QuSecure\rq~(Grant No. 13N14876) within the funding program Photonic Research Germany, the BMBF joint project \lq tubLAN Q.0\rq~(Grant No. 16KISQ087K), and by the Einstein Foundation via the Einstein Research Unit \lq Quantum Devices\rq~. C.S. gratefully acknowledges funding by the European Research Council via the project \lq unlimit-2D\rq~(ERC project 679288). The authors are grateful for helpful discussions as well as infrastructural support by Stephan Reitzenstein and technical assistance by Oliver Iff.

\section{Competing interests}
The authors declare no competing interests.

\section{Author contributions}
T.G. designed and built the QKD setup and the evaluation software used for the experiments. T.G. and M.v.H. operated the single-photon source, which was provided by C.A.S. and C.S.. T.G. performed the experiments and analyzed the data. T.G. and T.H. wrote the manuscript with input from all authors. T.H. conceived the experiment and supervised the project.

\section{Additional information}
Supplementary information is available.


\clearpage
\onecolumngrid

\renewcommand{\thepage}{S\arabic{page}} 
\renewcommand{\thesection}{S\arabic{section}}  
\renewcommand{\thetable}{S\arabic{table}}  
\renewcommand{\thefigure}{S\arabic{figure}}
\setcounter{figure}{0} 
\setcounter{page}{1}

\section{Supplementary Information}

\subsection{Photon arrival time distributions}
\begin{figure}[h]
	\centering\includegraphics[]{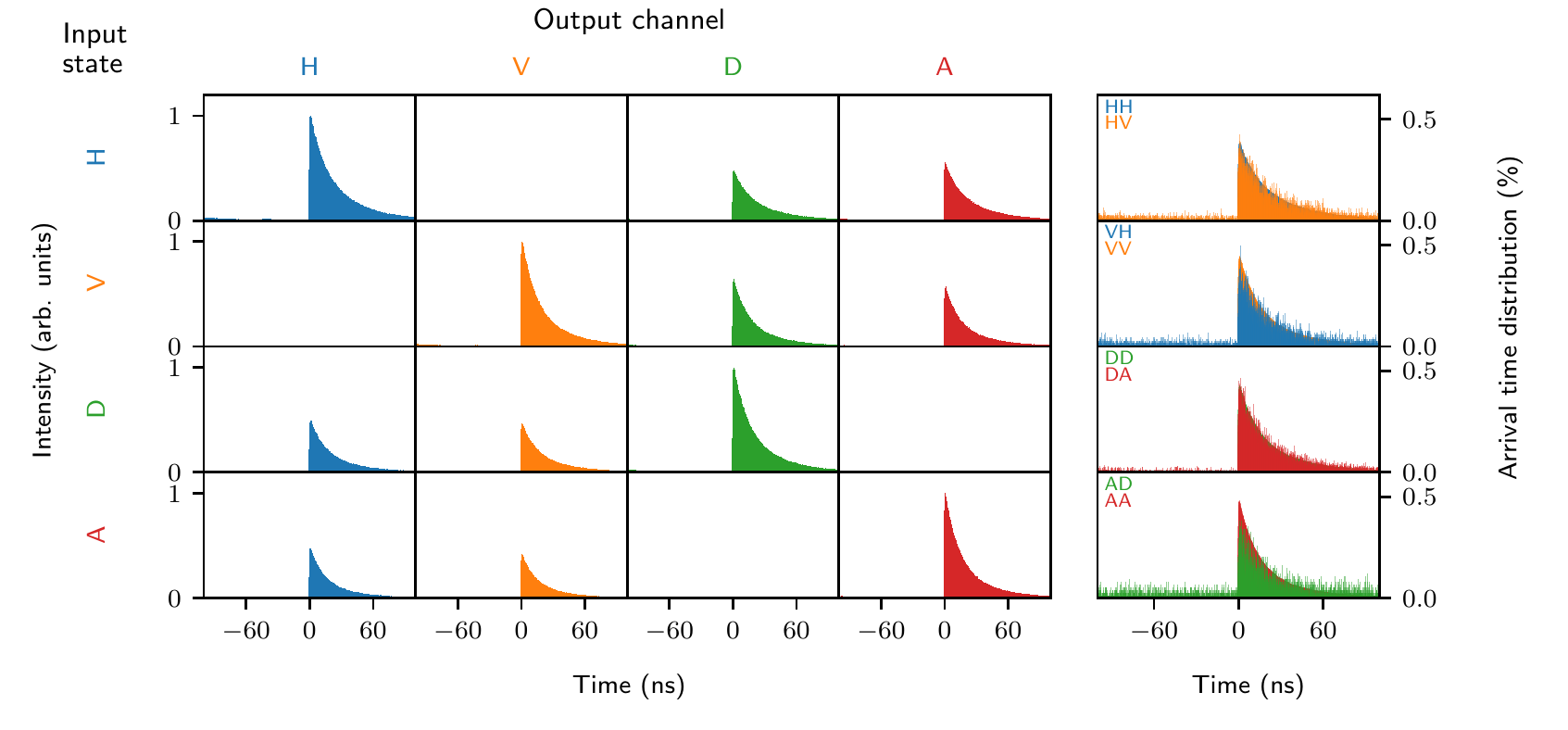}
	\caption{Photon arrival time distributions measured at the receiver module. For each of the four possible input polarizations H, V, D, and A, the time resolved photoluminescence is measured at each of the four corresponding output channels of the receiver module individually. The time-bin width is \SI{0.1}{\nano\second}. The measurement data in the left 4x4 matrix is normalized to the maximum of the respective input state. The right panel shows the two data sets for a given input polarization basis (e.g., HH and HV). Here, each distribution is normalized to the sum of all events registered in the respective channel (e.g., the HV distribution is normalized by all clicks registered in the V-channel).}
	\label{figS1}
\end{figure}

For investigating the performance of the receiver module, the receiver's response to all four possible input polarization states is investigated. We sequentially prepare single photon pulses in horizontal (H), vertical (V), diagonal (D), and anti-diagonal (A) polarization. The events are detected at all four detectors and individually correlated with the laser synchronization signal to obtain time-resolved measurements. These are summarized in Fig.~\ref{figS1} on the left-hand side in a 4x4 matrix. The rows correspond to the input polarization and the column to the respective detection channel. Histograms within one row are normalized to the maximum of the input polarization (e.g., the first row is normalized to the maximum of the HH distribution (blue)). The matrix in the main text Fig.~2\,b is calculated from the sum of each unnormalized matrix element in Fig.~\ref{figS1} divided by the acquisition time. Erroneous detection events can be made visible if the time-resolved measurement is renormalized to the arrival time probability distribution for both output channels in each basis, as depicted in the right panel of Fig.~\ref{figS1}.

\subsection{Effects of dynamic state preparation}

\begin{figure}
	\centering\includegraphics[]{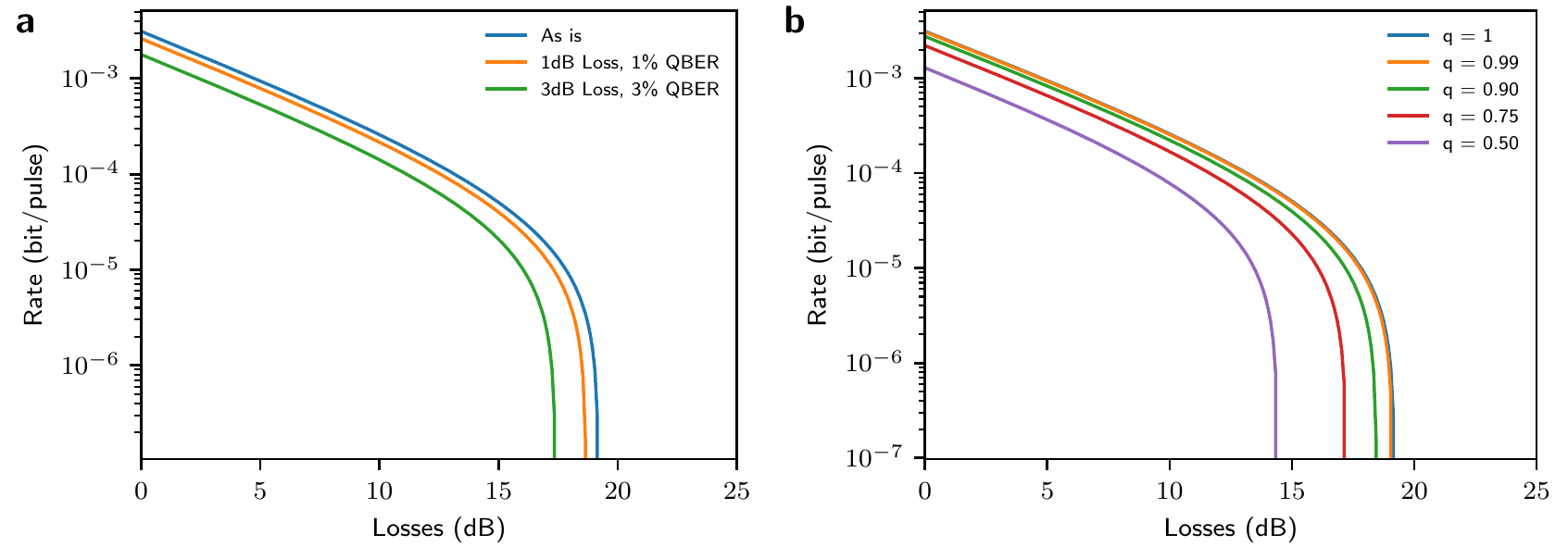}
	\caption{Impact of dynamic state preparation on the secret key rate vs. loss in full implementations of QKD. \textbf{a} Comparison of the original data show in the main text (\lq As is\rq~refers to the non-optimized case) and two realistic scenarios assuming dynamic encoding via an EOM introducing additional losses, i.e. reduced $\mu$, and increased QBER. \textbf{b} Impact of a finite state preparation qualities $q$.}
	\label{figS2}
\end{figure}
Our study demonstrates the feasability of quantum communication using single-photon sources based on 2D-TMDCs by using a QKD testbed with static polarization-state preparation. In full implementations, effects arising from a dynamic state preparation need to be considered in the performance and security analysis. Unlike WCP systems, however, the additional losses introduced by active modulation components like electro-optical modulators (EOMs) affect the achievable mean-photon number in the quantum channel. To gain a quantitative understanding, we consider two realistic scenarios assuming dynamic polarization-state preparation via free-space and fiber-based EOMs with typical transmission losses of \SI{1}{\decibel} and \SI{3}{\decibel}, respectively. The additional error EOMs contribute to the overal QBER (due to a finite extinction ratio) is typically in the range of \SI{1}{\percent} to \SI{3}{\percent}. Here, we assume the best case (\SI{1}{\decibel} loss, additional QBER \SI{1}{\percent}) and the worst case (\SI{3}{\decibel} loss, additional QBER \SI{3}{\percent}) scenarios for comparison. Figure~\ref{figS2}\,(a) compares the simulated rate-loss dependencies for both scenarios in comparison with the original data from our WSe$_2$ single-photon source (without temporal filtering) based on the parameters in Table~I of the main text. We find that the dynamic state preparation via EOMs noticeably affects the achievable secret key rate and tolerable loss, although the impact is moderate.
Another effect resulting from dynamic encoding is a finite state preparation fidelity or quality. While this effect was negligible in our proof-of-principle experiment using a crystal polarizer with high extinction ratio, the non-ideal state preparation will result in an additional contribution to the overall QBER in full implementations. In Fig.~\ref{figS2}\,b we investigated the effect of lower state preparation fidelity using the modified key rate equation from ref.~\cite{Tomamichel2012}:
\begin{equation}
	S_\infty=S_{\mathrm{sift}}\left[A\left(q-h\left(e/A\right)\right)-f_{\mathrm{EC}}h\left(e\right)\right],
\end{equation}
where $q$ denotes the state preparation quality. A high fidelity of \SI{90}{\percent} is desirable in full implementations for not loosing rate and tolerable losses.

\subsection{Temporal filtering}
\begin{figure}
	\centering\includegraphics[]{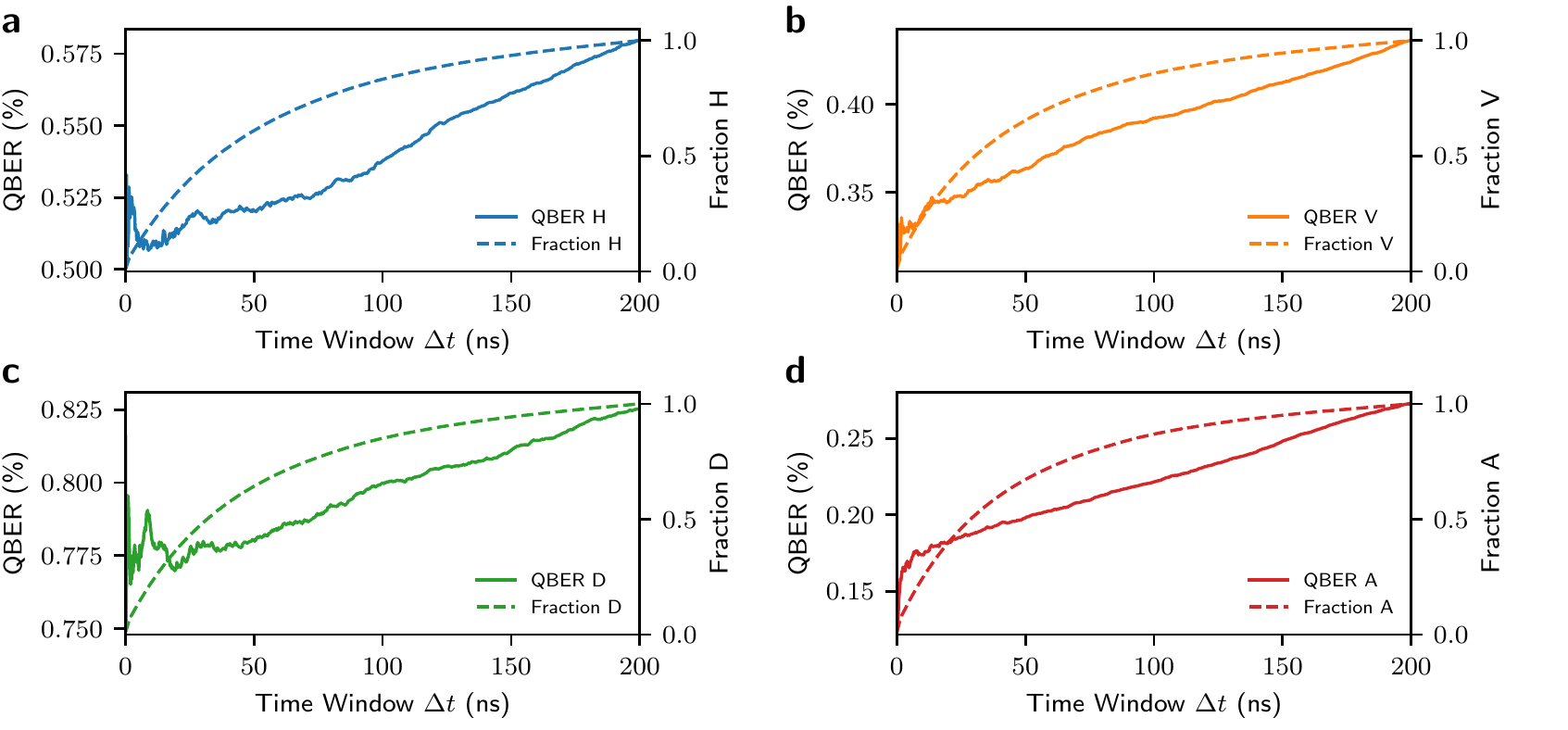}
	\caption{Temporal filtering with fixed $t_{\mathrm{c}} = 0$. \textbf{a} H input polarization. \textbf{b} V input polarization. \textbf{c} D input polarization. \textbf{d} A input polarization. From the photon arrival time distributions in Fig.~\ref{figS1} the QBERs and corresponding fractions of the sifted key are calculated depending on the width of the acceptance time window.}
	\label{figS3}
\end{figure}
In our study we exploit temporal filtering on the receiver side to optimize the achievable secure key rate and tolerable losses. For this purpose, an optimal trade-off between sifted key material and QBER is required. Figure~\ref{figS3} depicts the QBER and the sifted fraction for each detection channel in our four-state polarization analyzer as a function of the width $\Delta t$ of the acceptance time window ($t_{\mathrm{c}} = 0$). The plots are calculated from the experimental data in Fig.~\ref{figS1}. Reducing $\Delta t$, both the QBER and the sifted key fraction is reduced. The minimal QBER achieved is limited by optical imperfections in the experimental setup, as detector noise is effectively filtered by the acceptance time window.
\begin{figure}
	\centering\includegraphics[]{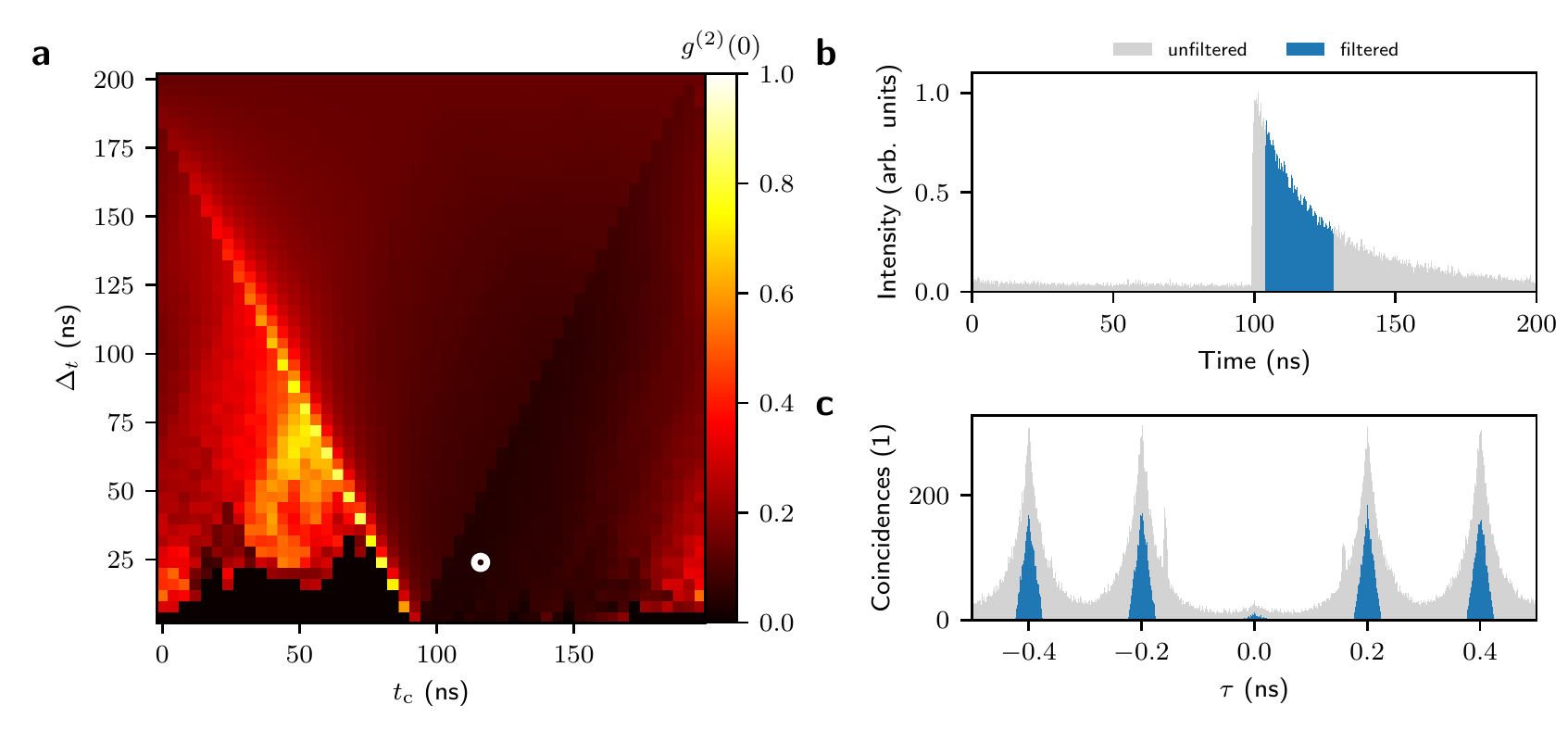}
	\caption{2D Temporal filtering of $g^{(2)}(0)$. \textbf{a} $g^{(2)}(0)$ as a function of the acceptance time window $\Delta t$ and $t_{\mathrm{c}}$. The $g^{(2)}(0) = 0.034\pm0.002$ from the main text is marked with a white circle ($\Delta t = \SI{24}{\nano\second}$, $t_{\mathrm{c}} = \SI{16}{\nano\second}$). \textbf{b} and \textbf{c} Photon arrival time distributions and autocorrelation histograms for the unfiltered (grey) and filtered (blue) case corresponding to the parameter set marked in \textbf{a}.}
	\label{figS4}
\end{figure} 
The temporal filtering does however not only affect the sifted key fraction and QBER, but also the $g^{(2)}(0)$. While in QKD one cannot benefit from temporal filtering of $g^{(2)}(0)$ without adding modifications to the protocol implementation (see main text for details), we make use of it to estimate contributions limiting the single-photon purity of our source. Figure~\ref{figS4}\,a shows the $g^{(2)}(0)$ of our $\mathrm{WSe}_2$ single-photon source as a function of the acceptance time window $\Delta t$ and $t_{\mathrm{c}}$. For this we used a subset of the experimental data (corresponding to the first \SI{15}{\minute}) from the measurement presented in Fig.~2\,d of the main text. For regions in black, i.e. zero, no reasonable $g^{(2)}(0)$ could be evaluated due to insufficient statistics. The minimum $g^{(2)}(0)$ is reached in the lower right quadrant of the 2D heatmap. In this region, signal from the exponential decay of the photon arrival time distribution of the quantum emitter is predominantly accepted, while residual background contributions of laser (left half, corresponds to rising edge of the photon arrival time distribution) and a large fraction of dark count contributions (lower left quadrant) is rejected. The white circle marks the value of $g^{(2)}(0) = 0.034\pm0.002$ at $\Delta t = \SI{24}{\nano\second}$, $t_{\mathrm{c}} = \SI{16}{\nano\second}$ we state in the main text, being in good agreement with our estimate on the spectral background of $0.04$ caused by uncorrelated background emission of the sample. This setting of the temporal filter results in a reasonable trade-off between the sifted key fraction (\SI{47}{\percent}) and the reduction of $g^{(2)}(0)$ from the unfiltered $g^{(2)}(0) = 0.139\pm0.002$. The photon arrival time distributions and corresponding autocorrelation histograms are shown in in Fig.~\ref{figS4}\,b and Fig.~\ref{figS4}\,c for the unfiltered (blue) and filtered (grey) case, respectively. While even lower $g^{(2)}(0)$ values can be achieved, tighter filtering leads to substantially reduced sifted key fractions. Comparing the two autocorrelation histograms in Fig.~\ref{figS4}\,c, also reveals that there is a significant difference between correlating the filtered detection events instead of using a linear background subtraction and only a certain interval around each peak of the unfiltered histogram for evaluating $g^{(2)}(0)$. 
\begin{figure}
	\centering\includegraphics[]{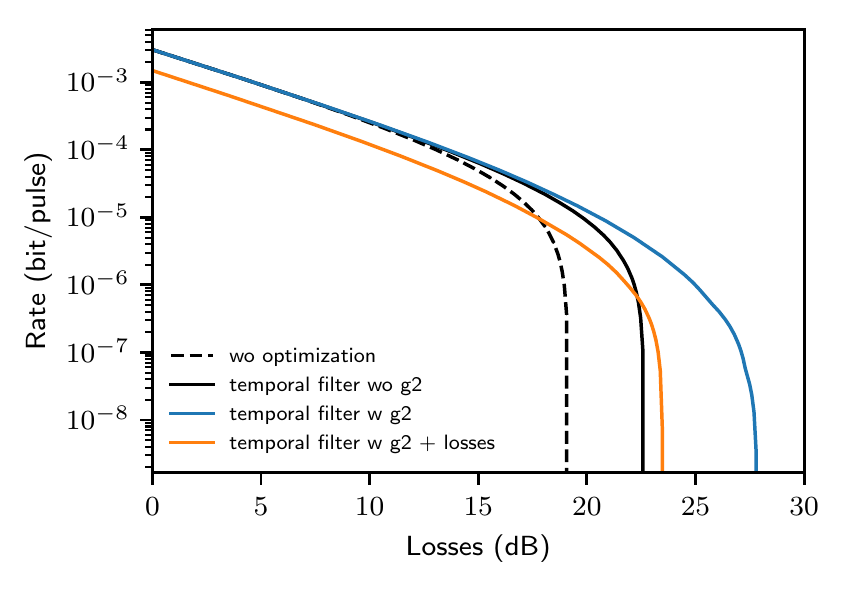}
	\caption{Impact of 2D temporal filtering of $g^{(2)}(0)$ on the secret key rate. The gain in tolerable loss compared to the original data from the main text (full and dashed black lines, with (w) and without (wo) temporal filtering, respectively) is only significant if additional losses of an EOM are neglected (blue line). In a more realistic scenario the additional \SI{3}{\decibel} loss of the modulator, required for active modulation on Alice's side, overcompensates the gain for our specific system (orange line).}
	\label{figS5}
\end{figure} 
For completeness, we also investigated how temporal filtering of $g^{(2)}(0)$ would affect the rate-loss dependencies in implementations with protocol extensions guaranteeing overall security, e.g. via active temporal filtering also on Alice side. Figure~\ref{figS5} compares the original data from the main text (full and dashed black lines, with (w) and without (wo) temporal filtering, respectively) with cases where the reduced $g^{(2)}(0)$ due to temporal filtering is taken into account (orange and blue lines). The comparison reveals a significant gain in tolerable loss only if additional losses of an EOM are neglected (blue line). Assuming \SI{3}{\decibel} of additional loss, the secure key rate is reduced while the tolerable loss becomes comparable to the optimized case without $g^{(2)}(0)$-filtering again. We assume \SI{3}{\decibel} loss of the modulator as the extinction ratio of fast fiber-based EOMs is typically higher ($\approx 1:10^4$) compared to the extinction ratio possible with free-space modulators ($\approx 1:10^2$). All solid lines result from an individual 2D optimization for each loss.

\subsection{Real-time photon statistics monitoring}
\begin{figure}
	\centering\includegraphics[]{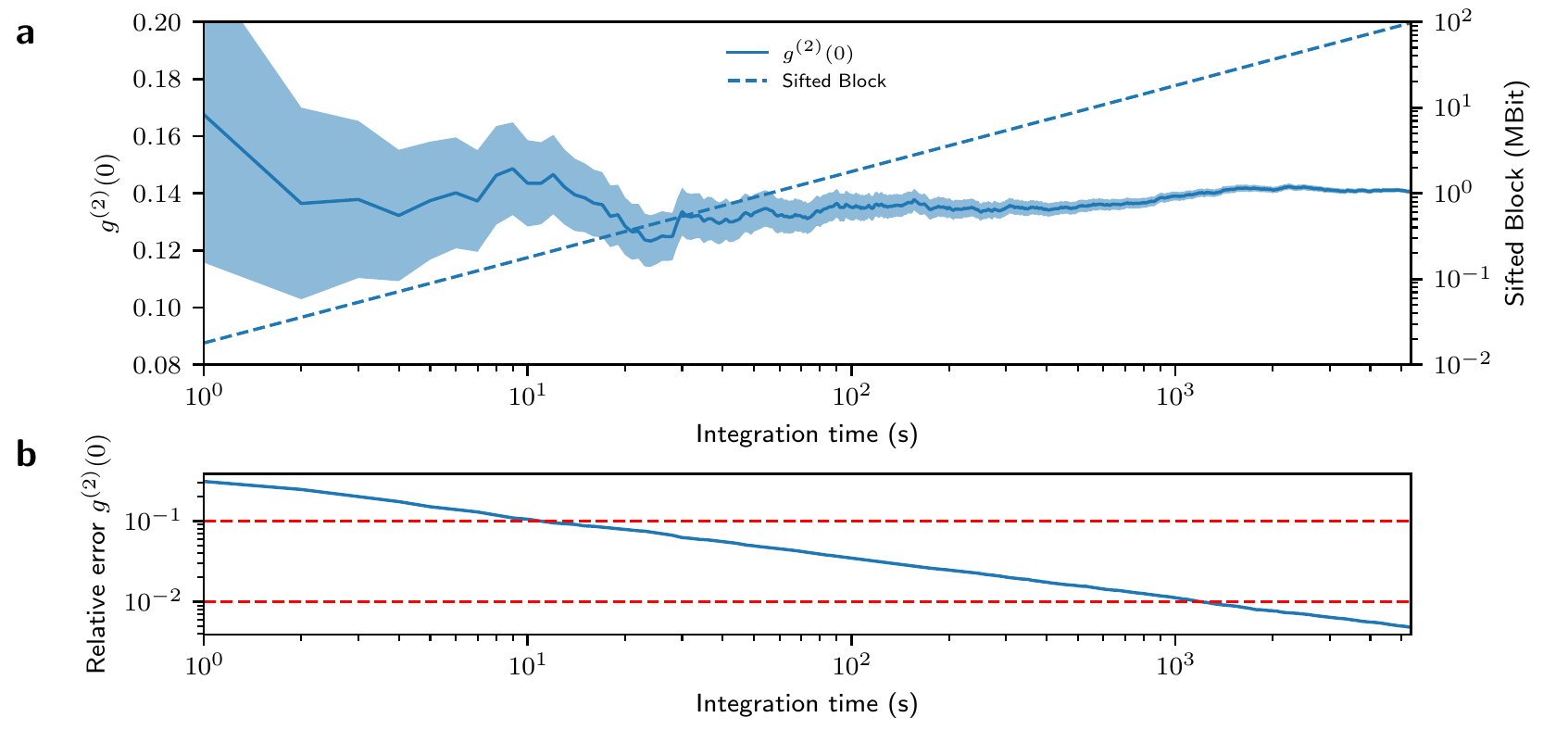}
	\caption{Antibunching investigation over integration time. \textbf{a} The antibunching $g^{(2)}(0)$ of the SPS is evaluated via correlating the detection events of all for detectors in overlapping blocks for different accumulation times compared with the corresponding length of the sifted block. Here, the same dataset as in Fig.~2\,d of the main text is used. \textbf{b} Relative error of $g^{(2)}(0)$ over the integration time. }
	\label{figS6}
\end{figure} 
For the real-time monitoring of the photon statistics, reasonable accumulation times need to be evaluated. In Fig.~2\,d of the main text, we choose \SI{10}{\second} and \SI{100}{\second}. The choice of these accumulation times is based on Fig.~\ref{figS6} depicting the antibunching $g^{(2)}(0)$ of the our $\mathrm{WSe}_2$ single-photon source as a function of the accumulation time. Here, the $g^{(2)}(0)$-value for accumulation times below \SI{103}{\second} is slightly lower even within the uncertainty bounds compared to the full measurement. This shows that a shorter acquisition time is suitable for taking changes in the emitter statistics into account. For \SI{10}{\second} accumulation time (\SI{0.178}{\mega bit} sifted block size), the relative error is about \SI{10}{\percent}, while an increase to \SI{100}{\second} (\SI{1.792}{\mega bit} sifted block size) reduces the relative error below \SI{5}{\percent}.

\end{document}